\documentclass{iopart}

\usepackage{graphicx}

\begin{document}

\title[Hyperbolic versus parabolic equation with fractional derivative]{Hyperbolic versus parabolic equation with fractional derivative to describe subdiffusion in a membrane system}

\author{Tadeusz Koszto{\l}owicz$^{1\dagger}$, Katarzyna D Lewandowska$^{2\ddagger}$ and Barbara G{\l}ogowska$^1$}
	
	\address{$^1$ Institute of Physics, University of Kielce,
         ul. \'Swi\c{e}tokrzyska 15, 25-406 Kielce, Poland.}

	\address{$^2$ Department of Physics and Biophysics, Medical University of
         Gda\'nsk, ul. D\c{e}binki 1, 80-211 Gda\'nsk, Poland.}
         
	\eads{\mailto{$^\dagger$tkoszt@pu.kielce.pl}, \mailto{$^\ddagger$kale@amg.gd.pl}}

\begin{abstract}
We use the parabolic and hyperbolic equation with fractional time derivative to describe the subdiffusion in a system with thin membrane. We find the Green's function and solutions of the equation for the system where the homogeneous solution is separated by a thin membrane from the pure solvent. The solutions were obtained for two boundary conditions where the ratio of the concentrations at the membrane surfaces does not change in time and where the flux flowing through the membrane is proportional to the concentration difference between membrane surfaces. We discuss the difference between the solutions for parabolic and hyperbolic subdiffusion equations obtained for both boundary conditions.
\end{abstract}


%\submitto{\JPA}

\maketitle

\section{Introduction}

The subdiffusion is usually defined as a process where the mean
square displacement of a particle $\left\langle \Delta
x^2\right\rangle$ is a power function of time
\begin{equation}\label{eq01}
    \left\langle\Delta x^2(t)\right\rangle=\frac{2D_\alpha}{\Gamma(1+\alpha)}t^\alpha ,
\end{equation}
where the subdiffusion parameter $\alpha$ is less than one
$(0<\alpha<1)$, $D_\alpha$ is the subdiffusion coefficient
measured in the units $m^2/s^\alpha$ \cite{mk}. The case of
$\alpha=1$ corresponds to the normal diffusion. The subdiffusion
is related to an infinitely long average time that a random walker
waits to make a finite jump. Then, its average displacement
square, which is observed in a finite time interval, is
dramatically suppressed. The subdiffusion occurs in systems with
complex internal structure such as gels or porous media. To
describe the subdiffusion the non-linear
differential equation of natural order derived on the base of Tsallis formalism \cite{tb,dhw} or the normal diffusion euqation with
diffusion coefficient which is assumed to be a power function of
time $D(t)=D_\alpha t^{\alpha-1}$ \cite{lm} were used. The Green's functions obtained
for these equations fulfill the relation (\ref{eq01}), but the
physical meaning of the assumptions providing to the equations is
not always clear. For example, it is difficult to explain the decreasing of diffusion coefficient in time in a homogeneous system. The subdiffusion linear equation with fractional time derivative, 
derived on the base of Continuous Time Random Walk formalism \cite{mk,compte96}, has not got such disadvantages.

The subdiffusion in a membrane system was recently studied
experimentally and theoretically. The motivation of the study is
that the understanding of subdiffusion in a membrane system can
help to model transport processes in systems so different as
living cells and membrane microfilters (see for example \cite{ra}). The system with a membrane
can also be used to measure subdiffusion parameters by comparing
theoretical and empirical concentration profiles of substances of
interest \cite{kdm}. To model a transport process in such a system the
parabolic subdiffusion equation was applied. However, the parabolic normal
diffusion and parabolic subdiffusion equations give the solutions which
posses the `unphysical' property. Namely, for spatially
unrestricted system the Green's function $G(x,t;x_0)$ (which is a
probability density of finding a particle in the position $x$
after time $t$ under condition that at the initial moment the particle
was located in $x_0$) have non-zero values for any $x$ and $t>0$.
This fact can be interpreted as an infinite speed propagation of
some particles. To avoid this `unphysical' property Cattaneo
proposed the hyperbolic normal diffusion equation based on
the assumption that the diffusion flux is delayed in time $\tau$ with
respect to the concentration gradient \cite{cattaneo}. The Green's function of the
equation is equal to zero for finite $x-x_0$, so the propagation
velocity of the particles is finite. In
phenomenological way the hyperbolic subdiffusion equation can be
derived by involving the fractional time derivative into a flux or
continuity equation. In the paper \cite{compte} there was noted
that the hyperbolic subdiffusion equation can be derived in three
different manners and the equations obtained are not equivalent to
each other.

The delaying effect of the flux with respect to the concentration gradient 
seems to be stronger in a membrane system
than in a homogeneous one, since the flux can be involved into
boundary conditions at the membrane. So, the delaying effect can
appear not only in the equation but also in the boundary
conditions. As far as we know, the hyperbolic subdiffusion
equation has not been applied to describe the subdiffusion in a
membrane system yet. In our paper we compare the solutions of the
parabolic and hyperbolic subdiffusion equation in a homogeneous
system and in a system with a thin membrane.

The problem of choosing a transport model in a membrane system is
more complicated since one of the boundary conditions at the
membrane is not set unambiguously. Two boundary
conditions which are not equivalent to each other were used. First of them
demands the constant ratio of the concentrations at the membrane
surfaces \cite{koszt1998,dkwm}, the second one assumes that the flux is proportional to
the concentration difference between membrane surfaces \cite{koszt2001a,koszt2001b}. The
qualitatively difference between them is manifested in the long
time limit because the concentrations calculated for second
boundary condition goes to a continuous function at the membrane,
unlike than for the first one.

In our paper we find the solutions of the hyperbolic equations for
a system with a thin membrane for two boundary conditions
mentioned above and compare them with the ones obtained from
parabolic equation. We consider the system where the thin membrane
separates a homogeneous solution from a pure solvent (we add that
such a system was often used in experimental studies \cite{kdm,dkwm,dsdoww,dwor2006}). In our study
we assume that the system is one-dimensional where the diffusion or
subdiffusion parameter as well as the parameter of membrane
permeability do not depend on time and concentration, the first
one is also independent of space variable.

The paper is organized as follows. In \sref{nde}  we present
the phenomenological derivation of the hyperbolic equation for normal
diffusion. We show plots of the Green's functions obtained for the
long times for the homogeneous system without a membrane. In \sref{subdifeq} we present the hyperbolic equation and the Green's functions for
the subdiffusive system. The boundary conditions at a thin membrane are derived in \sref{bcond}. Solutions of the hyperbolic equation for
the system where the homogeneous solution is separated by the
thin membrane from pure solvent are presented in \sref{sol}. To
illustrate our considerations the function obtained in sections
\ref{nde}, \ref{subdifeq} and \ref{sol} are shown  in several plots. Analyzing the plots we
discuss the properties of the solutions in \sref{sol}.

\section{Normal diffusion equation\label{nde}}

\subsection{Parabolic equation}

It is well known that the normal diffusion equation
\begin{equation}\label{eq0a}
    \frac{\partial C(x,t)}{\partial t}=D\frac{\partial^2 C(x,t)}{\partial x^2}
\end{equation}
with normal diffusion coefficient $D$ (measured in the units
$m^2/s$), can be derived phenomenologically by combining the first
Fick's law
\begin{equation}\label{eq0b}
        J(x,t)=-D\frac{\partial C(x,t)}{\partial x} ,
\end{equation}
and the continuity equation
\begin{equation}\label{eq3}
    \frac{\partial C(x,t)}{\partial t}=-\frac{\partial J(x,t)}{\partial x} .
\end{equation}
The Green's function is defined as a solution of the equation for the initial condition
\begin{equation}\label{eq5a}
    G(x,t;x_0)=\delta(x-x_0) ,
\end{equation}
and boundary conditions appropriate for considered system.
When the system is not spatially restricted, there is
\begin{equation}\label{eq0c}
    G(-\infty,t;x_0)=G(\infty,t;x_0)=0 ,
\end{equation}
and the Green's function reads
\begin{equation}\label{eq0d}
    G(x,t;x_0)=\frac{1}{2\sqrt{\pi Dt}}\exp\left(-\frac{(x-x_0)^2}{4Dt}\right) .
\end{equation}
The function (\ref{eq0d}) is different from zero for any $x$ and
$t>0$. Utilizing the probability interpretation of the Green's
function one concludes that some particles are transported with
the infinite speed propagation.

\subsection{Hyperbolic equation}

To ensure the finite velocity of the particle propagation one
assumes that the flux is delayed with respect to the concentration
gradient
\begin{equation}\label{eq1}
    J(x,t+\tau)=-D\frac{\partial C(x,t)}{\partial x} ,
\end{equation}
where $\tau$ is the delay time. Assuming that the parameter $\tau$ is sufficiently small, the left hand side of equation \eref{eq1} can be approximated by the first two terms of Taylor series with respect to $\tau$
\begin{equation}\label{eq2}
    J(x,t)+\tau\frac{\partial J(x,t)}{\partial t}=-D\frac{\partial C(x,t)}{\partial x} .
\end{equation}
Applying the operator $\partial/\partial x$ to equation (\ref{eq2})
and taking into considerations the continuity equation
(\ref{eq3}) one gets the hyperbolic diffusion equation
\begin{equation}\label{eq4}
    \tau\frac{\partial^2 C(x,t)}{\partial t^2}+\frac{\partial C(x,t)}{\partial t}=D\frac{\partial^2 C(x,t)}{\partial x^2} .
\end{equation}
We add that equation (\ref{eq4}) can be derived from
differential-difference equations with continuous time and
discrete space variable \cite{pottier}. The process can be
interpreted as a process with `minimal' memory which extends to
one time step more than in `ordinary' diffusion process described
by parabolic diffusion equation.

\Eref{eq4} ensures the finite propagation velocity of the particles
$v=\sqrt{D/\tau}$. In the limit $\tau\rightarrow 0$ we get
parabolic diffusion equation with infinite $v$. To solve equation
(\ref{eq4}) we must take two initial conditions. Let us assume
that one of them is
\begin{equation}\label{eq5}
    \left.\frac{\partial C(x,t)}{\partial t}\right|_{t=0}=0 ,
\end{equation}
what means that at an initial moment the concentration does not
aim at its change and is effectively changed after time $\tau$ since the
particle flux is not generated before this time. The second boundary
condition reads as $C(x,0)=f(x)$.

\subsection{Green's function}

We obtain the Green's function for equation (\ref{eq4}) solving it by
means of the Laplace $L[f(t)]=\hat{f}(s)=\int_0^\infty
f(t)\exp(-st)dt$ and Fourier
$F[g(x)]=\hat{g}(k)=\int_{-\infty}^\infty g(x)\exp(ikx)dx$
transforms method for the initial conditions (\ref{eq5a}) (with $x_0=0$) and (\ref{eq5}). After simple calculations we get the
Green's function in terms of Lapalce and Fourier transforms
\begin{equation}\label{eq6}
    \hat{G}(k,s;0)=\frac{1+\tau s}{s+\tau s^2+Dk^2} .
\end{equation}
The inverse Fourier transform of equation (\ref{eq6}) reads
\begin{equation}\label{eq7}
    \hat{G}(x,s;0)=\frac{\sqrt{1+\tau
    s}}{2\sqrt{Ds}}\exp\left(-\frac{|x|\sqrt{s}}{\sqrt{D}}\sqrt{1+\tau s}\right) .
\end{equation}
The hyperbolic equation was derived on the assumption that we omit the terms which iclude the parameter $\tau^k$, $k>1$, in the Taylor series of the flux
(see equation (\ref{eq2})). We find
similar approximation of equation (\ref{eq7}), namely
\begin{equation}\label{eq8}
     \hat{G}(x,s;0)=\frac{1}{2\sqrt{Ds}}\left(1+\frac{\tau\sqrt{s}}{2}-\frac{|x|s}{2\sqrt{D}}\right)
     \exp\left(-\frac{|x|\sqrt{s}}{\sqrt{D}}\right) .
\end{equation}
The inverse Laplace transform of equation (\ref{eq8}) is
\begin{equation}\label{eq9}
\fl   G(x,t;0)=\frac{1}{2\sqrt{\pi Dt}}\left(1+\frac{|x|\tau}{4t\sqrt{D}}\right)\exp{\left(-\frac{x^2}{4Dt}\right)}
    -\frac{|x|\tau}{4D}f_{1/2,1/2}\left(t;\frac{|x|}{\sqrt{D}}\right) ,
\end{equation}
where the function $f$ is defined as \cite{koszt2004}
\begin{displaymath}
    f_{\nu,\beta}(t;a)\equiv L^{-1}\left[s^\nu\exp\left(-as^\beta\right)\right] ,
\end{displaymath}
for $a,\beta>0$.
This function can be expressed by the Fox function $H$ and reads
\begin{eqnarray}\label{eq9a}
    f_{\nu,\beta}(t;a)&=&\frac{1}{\beta a^{(1+\nu)/\beta}}H^{1 0}_{1 1}\left(\left.\frac{a^{1/\beta}}{t}\right|
    			\begin{array}{cc}
             1 & 1 \\
             (1+\nu)/\beta & 1/\beta
           \end{array}
\right) \nonumber\\
&=&\frac{1}{t^{1+\nu}}\sum_{k=0}^{\infty} \frac{1}{k!\Gamma(-k\beta-\nu)}\left(-\frac{a}{t^\beta}\right)^k .
\end{eqnarray}

\begin{figure}[h]
		\centering
    \includegraphics{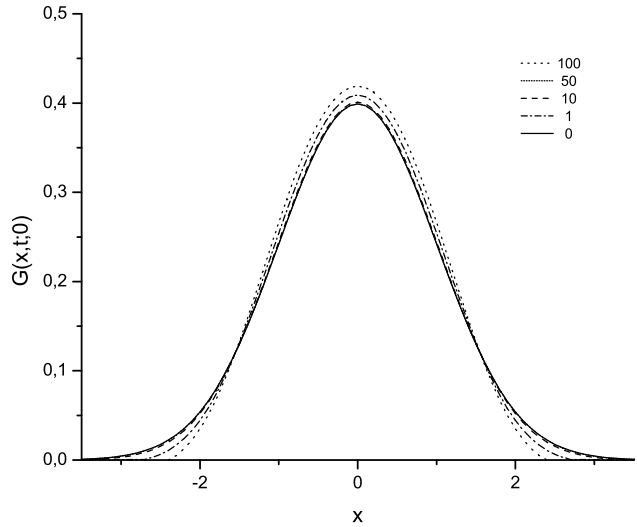}
    \caption{The plots of the normal diffusion Green's functions for different values of parameter $\tau$ given in the legend, here $t=500$, $D=10^{-3}$.\label{fig:Fig1}}
\end{figure}

The plots of function (\ref{eq9}) are presented in
\fref{fig:Fig1} for different values of the parameter $\tau$.
As we can see, only relatively large values of $\tau$ make the
noticeable difference between the Green's functions obtained for
the parabolic equation (represented by the solutions for $\tau=0$) and the hyperbolic one.

\section{Subdiffusion equation\label{subdifeq}}

\subsection{Parabolic equation}

The hyperbolic subdiffusion equation can be derived by analogy with
the derivation of the parabolic one. There are few ways
to find the parabolic subdiffusion equation in phenomenological
way. In the following we consider two of them which are natural generalization of the derivation of the parabolic normal diffusion equation. In the first one it is assumed that the subdiffusive flux
reads as
\begin{equation}\label{eq10f}
    J(x,t)=-D_\alpha\frac{\partial_{\rm RL}^{1-\alpha}}{\partial t^{1-\alpha}}\frac{\partial C(x,t)}{\partial x},
\end{equation}
where $\partial_{\rm RL}^\alpha/\partial t^\alpha$ denotes the Riemann-Liouville fractional time derivative defined as \cite{os,pod} (here $\alpha>0$)
\begin{equation}\label{eq10a}
    \frac{d_{\rm RL}^{-\alpha}f(t)}{dt^{-\alpha}}=\frac{1}{\Gamma(\alpha)}\int^{t}_{0}(t-u)^{\alpha-1}f(u)du ,
\end{equation}
and
\begin{equation}\label{eq10b}
    \frac{d_{\rm RL}^{\alpha}f(t)}{dt^{\alpha}}=\frac{d^n}{dt^n}\frac{d_{\rm RL}^{\alpha-n}f(t)}{dt^{\alpha-n}} ,
\end{equation}
where $n$ is the lowest natural number that fulfills $n\geq\alpha$.
The Laplace transform of the Riemann-Liouville derivative is
\begin{equation}\label{eq10c}
    L\left[\frac{d^\alpha_{\rm RL}
    f(t)}{dt^\alpha}\right]=s^\alpha\hat{f}(s)-\left.
    \sum_{k=0}^{n-1}s^k\frac{d^{\alpha-k-1}_{\rm RL}f(t)}{dt^{\alpha-k-1}}\right|_{t=0} ,
\end{equation}
where $n-1\leq\alpha<n$. Combining equation (\ref{eq10f}) with equation
(\ref{eq3}) one gets the parabolic subdiffusion equation \cite{mk,compte96}
\begin{equation}\label{eq10g}
    \frac{\partial C(x,t)}{\partial t}=D_\alpha\frac{\partial_{\rm RL}^{1-\alpha}}{\partial
    t^{1-\alpha}}\frac{\partial^2 C(x,t)}{\partial x^2} .
\end{equation}
In general, to solve the differential equation with the fractional Riemann-Liouville
derivative by means of the Laplace transform method, one should
fix the initial condition for time derivatives of the fractional
negative order (see equation (\ref{eq10c})), what is beyond of physical
interpretation. However, this remark does not concern the
subdiffusion equation (\ref{eq10g}) since for a limited function
there is (see Appendix)
\begin{equation}\label{eq10s}
\left.\frac{\partial_{\rm RL}^{\alpha-1}C(x,t)}{\partial t^{\alpha-1}}\right|_{t=0}=0 ,
\end{equation}
when $0<\alpha<1$.
The Laplace and Fourier transforms of (\ref{eq10g}) is
\begin{equation}\label{eq10h}
    s\hat{C}(k,s)-F[C(x,0)]=-D_\alpha s^{1-\alpha}k^2\hat{C}(k,s) .
\end{equation}
For the Green's function with the initial condition (\ref{eq5a}) we get
$F[C(x,0)]=1$, what leads to the form of equation (\ref{eq10h})
obtained from the Continuous Time Random Walk formalism
\cite{mk}.

Another scenario provides to the subdiffusion equation consists in
changing the time derivative of natural order to the fractional Caputo one in the continuity
equation (\ref{eq3}) according to the formula
$\partial/\partial t\rightarrow\theta\partial^\alpha_{\rm C}/\partial t^\alpha$ ,
where $\theta$ is a parameter which is involved to achieve an
appropriate physical units. The Caputo fractional derivative is defined by the relation \cite{pod}
\begin{equation}\label{eq10d}
\frac{d_{\rm C}^{\alpha}f(t)}{dt^{\alpha}}=\frac{1}{\Gamma(n-\alpha)}\int^{t}_{0}(t-u)^{\alpha-1}\frac{d^n}{dt^n}f(u)du ,
\end{equation}
and its Laplace transform reads as
\begin{equation}\label{eq10e}   L\left[\frac{d_{\rm C}^\alpha f(t)}{dt^\alpha}\right]=s^\alpha\hat{f}(s)-\left.\sum_{k=0}^{n-1}s^{\alpha-k-1}\frac{d^k f(t)}{dt^k}\right|_{t=0} ,
\end{equation}
where $n-1\leq\alpha<n$.
Thus, we get
\begin{equation}\label{eq10k}
    \theta\frac{\partial^\alpha_{\rm C} C(x,t)}{\partial t^\alpha}=-\frac{\partial J(x,t)}{\partial x} .
\end{equation}
In the following we take
\begin{equation}\label{eq10l}
    \theta=\frac{D}{D_\alpha} .
\end{equation}
Combining (\ref{eq0b}), (\ref{eq10k}) and (\ref{eq10l}) one gets the subdiffusion equation
\begin{equation}\label{eq10j}
    \frac{\partial_{\rm C}^\alpha C(x,t)}{\partial t^\alpha}=D_\alpha\frac{\partial^2 C(x,t)}{\partial x^2} .
\end{equation}
\Eref{eq10j} is equivalent to equation (\ref{eq10g}) since their Laplace and Fourier transforms 
are expressed by (\ref{eq10h}).

\subsection{Hyperbolic equation}

The hyperbolic subdiffusion equation can be obtained by
introducing the time derivative of fractional order to equations
(\ref{eq3}) or (\ref{eq2}). As was noticed in the paper
\cite{compte}, where the
Riemman-Liouville fractional derivative only was taken into considerations, it can be done in three different manners. Unlike in \cite{compte}, we involve Caputo fractional derivative into
the continuity equation (\ref{eq3}). We assume that the flux is given as follow
\begin{equation}\label{eq10}
    J(x,t+\tau)=-D_\alpha\frac{\partial^{1-\alpha}_{\rm RL}}{\partial t^{1-\alpha}}\frac{\partial C(x,t)}{\partial x} .
\end{equation}
Similarly to the previous case, let us approximate the left hand of equation (\ref{eq10}) for $\tau\ll t$ by the first two terms of Taylor series with respect to $\tau$
\begin{equation}\label{eq11}
        J(x,t)+\tau\frac{\partial J(x,t)}{\partial t}=-D_\alpha\frac{\partial^{1-\alpha}_{\rm RL}}{\partial t^{1-\alpha}}\frac{\partial C(x,t)}{\partial x} .
\end{equation}
From equation (\ref{eq11}) and equation (\ref{eq3}) we get the
hyperbolic subdiffusion equation
\begin{equation}\label{eq12}
    \tau\frac{\partial^2 C(x,t)}{\partial t^2}+\frac{\partial C(x,t)}{\partial t}=D_\alpha\frac{\partial^{1-\alpha}_{\rm RL}}{\partial t^{1-\alpha}}\frac{\partial^2 C(x,t)}{\partial x^2} .
\end{equation}
Let us note that from equations (\ref{eq2}), (\ref{eq10k}) and
(\ref{eq10l}) we get the hyperbolic subdiffusion
equation with Caputo fractional derivatives
\begin{equation}\label{eq10m}
    \tau\frac{\partial^{1+\alpha}_{\rm C} C(x,t)}{\partial t^{1+\alpha}}+\frac{\partial^\alpha_{\rm C} C(x,t)}{\partial t^\alpha}=D_\alpha\frac{\partial^2 C(x,t)}{\partial x^2} .
\end{equation}
\Eref{eq10m} is fully equivalent to equation (\ref{eq12}) since
the Laplace and Fourier transforms of the equations are the same.

\subsection{Green's function}

As previous, we take the initial conditions (\ref{eq5a}) (for $x_0=0$) and
(\ref{eq5}) to solve equation (\ref{eq12}). After
calculations we get
\begin{equation}\label{eq13}
    \hat{G}(k,s;0)=\frac{1+\tau s}{s+\tau s^2+D_\alpha s^{1-\alpha}k^2} ,
\end{equation}
the inverse Fourier transform of equation (\ref{eq13}) is
\begin{equation}\label{eq14}
    \hat{G}(x,s;0)=\frac{\sqrt{1+\tau s}}{2\sqrt{D_\alpha}
    s^{1-\alpha/2}}\exp\left(-\frac{s^{\alpha/2}|x|\sqrt{s}}{\sqrt{D_\alpha}}\sqrt{1+\tau s}\right) .
\end{equation}

The hyperbolic equation was derived under the assumption that we take
into account linear terms in the Taylor series of the flux
(see equation (\ref{eq2})). Let us perform similar approximation for equation (\ref{eq7}), what gives
\begin{equation}\label{eq15}
    \hat{G}(x,s;0)=\frac{1}{2\sqrt{D_\alpha}s^{1-\alpha/2}}\left(1+\frac{\tau s^{\alpha/2}}{2}
    -\frac{|x|s^{\alpha}}{2\sqrt{D_\alpha}}\right)\exp\left(-\frac{|x|s^{\alpha/2}}{\sqrt{D_\alpha}}\right) .
\end{equation}
The inverse Laplace transform of equation (\ref{eq15}) is
\begin{eqnarray}\label{eq16}
\fl G(x,t;0)=\frac{1}{2\sqrt{ D_\alpha}}\left[f_{\alpha/2-1,\alpha/2}\left(t;\frac{|x|}{\sqrt{D_\alpha}}\right)
    +\frac{\tau}{2}f_{\alpha/2,\alpha/2}\left(t;\frac{|x|}{\sqrt{D_\alpha}}\right)\right. \nonumber\\
 \left.-\frac{|x|\tau}{2\sqrt{D_\alpha}}f_{\alpha,\alpha/2}\left(t;\frac{|x|}{\sqrt{D_\alpha}}\right)\right].
\end{eqnarray}

\begin{figure}[h]
		\centering
    \includegraphics{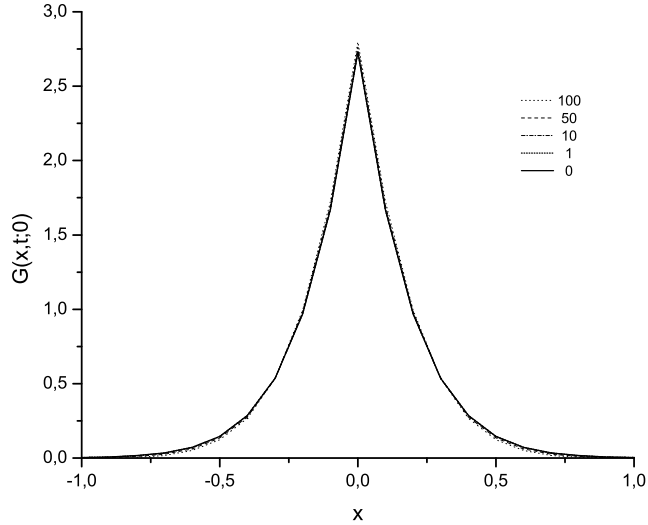}
    \caption{Hyperbolic subdiffusion. The plots of the Green's functions for different values of $\tau$, here $t=500$, $D_\alpha=10^{-3}$, $\alpha=0.5$.\label{fig:Fig2}}
\end{figure}

\begin{figure}[h]
		\centering
    \includegraphics{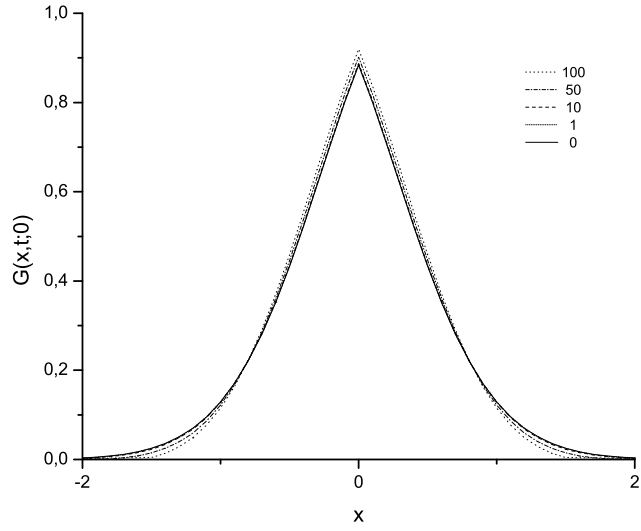}
    \caption{The description as in \fref{fig:Fig2} but for $\alpha=0.8$.\label{fig:Fig3}}
\end{figure}

\begin{figure}[h]
		\centering
    \includegraphics{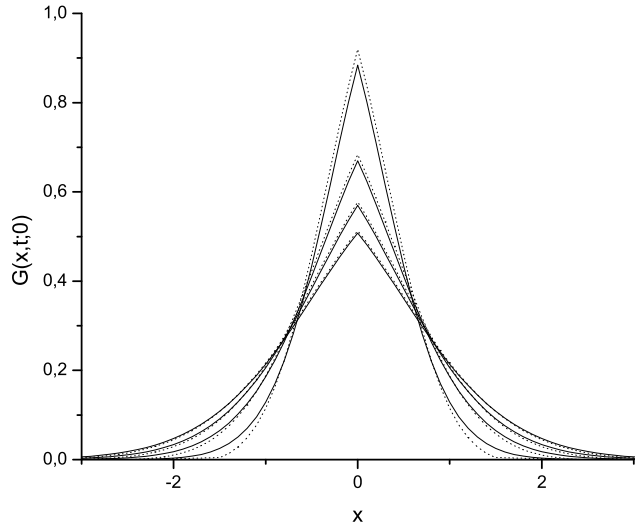}
    \caption{The Green's functions for $t=500, 1000, 1500, 2000$, here $D_\alpha=10^{-3}$, $\alpha=0.8$, the dashed lines correspond to $\tau=100$, continuous ones represent the Green's functions with $\tau=0$.\label{fig:Fig4}}
\end{figure}

The plots of the Green's functions (\ref{eq16}) are presented in figures \ref{fig:Fig2}-\ref{fig:Fig4}. Contrary to the normal diffusion case, the effect of delaying is hardly observed in the considered cases.

\section{Boundary conditions at thin membrane\label{bcond}}

We denote the concentration and flux in the region $x<x_m$ as $C_1$
and $J_1$, and in the region $x>x_m$ as $C_2$ and $J_2$,
respectively, where $x_m$ is the membrane position. Since the equation is of the second order with
respect to the space variable, we need two boundary conditions in
each of the region. Two boundary conditions demand finiteness of
the solutions at $x\rightarrow -\infty$ and $x\rightarrow \infty$,
two other ones are fixed at the membrane. The first of them is rather
obvious and it assumes the continuity of the flux at the membrane
\begin{equation}\label{eq18}
    J_1(x_m^-,t)=J_2(x_m^+,t)\equiv J(x_m,t) .
\end{equation}
However, the problem of fixing the second boundary condition at
the thin membrane is not unambiguously solved. The missing boundary condition at the
membrane has been chosen in two ways. The first one demands the
constant ratio of the concentrations at two opposite sides of the
membrane \cite{kdm,koszt1998,dkwm}
    \begin{equation}\label{5}
C_2(x^+_m,t)/C_1(x^-_m,t)=\gamma=const.
    \end{equation}
In the second one the flux flowing
through the membrane is proportional to the difference of the
concentrations at the opposite sides of the membrane \cite{kdm,koszt2001a,koszt2001b}
    \begin{equation}\label{5a}
J(x_m,t)=\lambda[C_2(x^-_m,t)-C_1(x^+_m,t)] ,
    \end{equation}
where $\lambda$ is the membrane permeability coefficient.
Below we consider the possibility of application of the boundary
conditions (\ref{5}) or (\ref{5a}) for the case of the system
described by the hyperbolic normal diffusion or hyperbolic subdiffusion equation. 

\subsection{Constant concentration ratio at the membrane}

In the Smoluchowski's papers (see for example \cite{smol}) it was derived the boundary condition at the fully reflecting wall. Let the wall is placed at $x_m$ and the system occupies
the interval $(-\infty,x_m)$. Smoluchowski's original approach
utilized the assumptions that the amount of the substance in the
system does not change in time
    \begin{equation}\label{eq5c}
\frac{\partial}{\partial t}\int^{x_m}_{-\infty}C(x,t)dx=0 ,
    \end{equation}
and the flux vanishes at $-\infty$. Integrating the continuity
equation (\ref{eq3}) over the interval $(-\infty, x_m)$ and using
the above assumptions one gets
    \begin{equation}\label{eq5d}
J(x_m,t)=0 .
    \end{equation}
Since equation (\ref{eq3}) works in hyperbolic subdiffusion equation case, we choose the boundary condition (\ref{eq5d}) at the fully reflecting
wall for the system described by the hyperbolic equation.

Chandrasekhar used the method of images to derive the Green's
function for this system \cite{chan}. The Green's function can be interpreted
as a concentration of large number of particles $N$ (divided by
$N$); the particles are located at point $x_0$ at the initial
moment $t=0$. So, the Green's function can be treated as an
instantaneous particle source (IPS) normalized to $1$. Within the
method one replaces the wall by additional IPS in such a manner
that the concentration behaves exactly as in the system with the
wall. Vanishing of the flux at the reflecting wall is achieved
when one replaces the wall by the IPS located symmetrically to the
initial point $x_0$ with respect to the wall. Then the
instantaneous particle sources create fluxes of particles flowing
in opposite directions, which reduce each other at the point
$x_m$. Thus, one finds (for $x<x_m$ and $x_0<x_m$)
        \begin{equation}\label{6}
G(x,t;x_0)=G_0(x,t;x_0)+G_0(x,t;2x_m-x_0) ,
        \end{equation}
where $G_0$ denotes the Green's function for a homogeneous system
(without the wall). Let us note that the Green's function equation
(\ref{6}) leads to equation (\ref{eq5d}) in any system where the
flux fulfills the following relation
    \begin{equation}\label{jpgs}
J \sim \frac{\partial C}{\partial x} .
    \end{equation}
In the papers \cite{koszt1998} the method of images was generalized to the
system with a partially permeable wall. Since the Green's function (\ref{6}) works in the system with fully reflecting wall, where a transport is described by hyperbolic diffusion or subdiffusion equation, similar generalization can be performed in such a system with thin membrane.

In the system with thin
membrane the particles can pass trough the membrane in both
directions many times. Let us assume that the membrane is
symmetric and the probabilities of passing through it do not
depend on the direction of particle's motion. To take into account
selective properties of the membrane, we `weaken' the additional
IPS located at $2x_m-x_0$ by factor $\sigma$, which gives
        \begin{equation}\label{7}
G(x,t;x_0)=G_0(x,t;x_0)+\sigma G_0(x,t;2x_m-x_0) .
        \end{equation}
Assuming that the flux is continuous at the membrane, we get for $x>x_m$ and $x_0<x_m$
        \begin{equation}\label{8}
G(x,t;x_0)=(1-\sigma)G_0(x,t;x_0).
        \end{equation}
The functions (\ref{7}) and (\ref{8}) fulfill the boundary
condition (\ref{5}) with $\lambda=(1-\sigma)/(1+\sigma)$. To
interpret the parameter $\sigma$ let us note that the probability
of finding a particle (starting from $x_0$, where $x_0<x_m$) in
the region $x>x_m$ is equal to
$P_\sigma=(1-\sigma)\int^{\infty}_{x_m}G_0(x,t;x_0)dx$ for the
membrane system, whereas the probability of finding the particle
in this region for system without the membrane is equal to
$P_0=\int^{\infty}_{x_m}G_0(x,t;x_0)dx$. Comparing the above
equations we obtain $\sigma=1-P_\sigma/P_0$, so the parameter
$\sigma$ can be interpreted as a probability of finding the
particle in the region $x<x_m$ under condition that in the system
with removed membrane the particle will be in the region $x>x_m$.
In other words, $\sigma$ is a conditional probability of stopping
the particle by the membrane in unit time under condition, that in
the similar system with no membrane this particle pass the
position $x_m$. Thus, $\sigma$ is the parameter controlling the
reflection of particles by the membrane, the parameter $1-\sigma$
is the parameter of membrane permeability. The boundary condition
(\ref{5}) has simple physical interpretation: {\it if $N$
diffusing particles are going to pass through the wall in unit
time, then $\sigma N$ of them will be stopped by the wall whereas
$(1-\sigma)N$ pass through, where $\sigma
=(1-\lambda)/(1+\lambda)$}.

\subsection{Radiation boundary condition}

For the parabolic normal diffusion equation the radiation boundary condition (\ref{5a}) was derived from the model with discrete space variable \cite{koszt2001a} as well as for the considerations
performed in a phase space where the diffusion is described by the
Klein-Kramers equation \cite{koszt2001b}.
\Eref{5a} can be interpreted as the natural continuation of
the Fick equation applied to the membrane. According to equation
(\ref{eq1}), we can generalize equation (\ref{5a}) as follows
\begin{equation}\label{eq27a}
    J(x_m,t+\tau)=\lambda[C_1(x_m^-,t)-C_2(x_m^+,t)] .
\end{equation}
In the following we will see that the boundary conditions (\ref{5}) and (\ref{eq27a}) are
not equivalent to each other. 

\section{Solutions of hyperbolic subdiffusion equation for a membrane system\label{sol}}

Let us assume that the thin membrane is located at $x_m=0$. We
choose the initial condition as
\begin{equation}\label{eq21}
    C(x,0)=\left\{ \begin{array}{cc}
             C_{0}, & x<0 \\
             0, & x>0 .
           \end{array} \right.
\end{equation}
The boundary conditions demand finiteness of the solutions at infinity
\begin{equation}\label{eq17}
    C_1(-\infty,t)=C_0,\; C_2(\infty,t)=0 .
\end{equation}
The first boundary condition at the membrane (\ref{eq18}) ensures that the flux is continuous at the one,
the second boundary condition at the membrane we take in general form
\begin{equation}\label{eq19}
    b_1C_1(0^-,t)+b_2C_2(0^+,t)+b_3J(0,t+\tau)=0 .
\end{equation}
We note that the Laplace transform of the flux reads
\begin{equation}\label{eq20}
    \hat{J}(x,s)=-D_\alpha\frac{s^{1-\alpha}}{1+\tau s}\frac{d\hat{C}(x,s)}{dx} .
\end{equation}
The Laplace transforms of solutions for boundary conditions
(\ref{eq18}), (\ref{eq17}), (\ref{eq19}) and initial condition
(\ref{eq21}) are as follows
\begin{equation}\label{eq22}
\fl\hat{C}_1(x,s)=\frac{C_0}{s}\left[1-\frac{b_1}{b_1-b_2-b_3\sqrt{D_\alpha}s^{1-\alpha/2}/\sqrt{1+\tau s}} \exp\left(x\sqrt{\frac{(1+\tau s)s^\alpha}{D_\alpha}}\right)\right] ,
\end{equation}
\begin{equation}\label{eq23}
 \fl  \hat{C}_2(x,s)=\frac{C_0}{s}\frac{b_1}{b_1-b_2-b_3\sqrt{D_\alpha}s^{1-\alpha/2}/\sqrt{1+\tau s}}
    \exp\left(-x\sqrt{\frac{(1+\tau s)s^\alpha}{D_\alpha}}\right) .
\end{equation}
Below we find the solutions for two boundary conditions described in \sref{bcond}.

\subsection{Constant ratio of the solutions at the membrane}

\begin{figure}[h]
		\centering
    \includegraphics{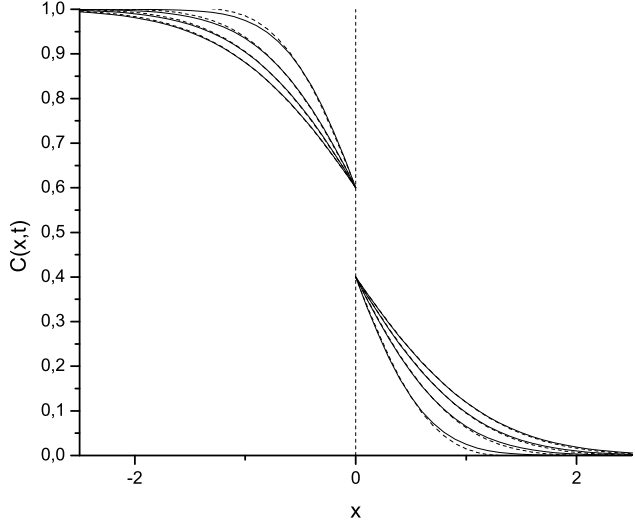}
    \caption{The solutions calculated for the boundary condition (\ref{5})
    with $\gamma=1.5$, $\alpha=0.9$, $D_\alpha=5\times 10^{-4}$ for $t=500, 1000, 1500, 2000$. Vertical lines represent the membrane, dashed lines correspond to $\tau=100$, continuous ones correspond to $\tau=0$.\label{fig:Fig5}}
\end{figure}

Putting $b_1>0$, $b_2<0$ and $b_3=0$ in (\ref{eq19}) we get equation (\ref{5}) with $\gamma=-b_2/b_1$.
The solutions are as follows
\begin{equation}\label{eq25}
    C_1(x,t)=C_0\left[1-\sigma f_{-1,\alpha/2}\left(t;\frac{-x}{\sqrt{D_\alpha}}\right)
-\sigma\frac{x\tau}{2\sqrt{D_\alpha}}f_{\alpha/2,\alpha/2}\left(t;\frac{-x}{\sqrt{D_\alpha}}\right)\right],
\end{equation}
\begin{equation}\label{eq26}
    C_2(x,t)=C_0\sigma\left[f_{-1,\alpha/2}\left(t;\frac{x}{\sqrt{D_\alpha}}\right)
-\frac{x\tau}{2\sqrt{D_\alpha}}f_{\alpha/2,\alpha/2}\left(t;\frac{x}{\sqrt{D_\alpha}}\right)\right],
\end{equation}
where $\sigma=1/(1+\gamma)$. The plots of functions (\ref{eq25})
and (\ref{eq26}) are presented in \fref{fig:Fig5}.

As we can see, the differences between the solutions obtained for
the parabolic subdiffusion equation are very close to the solution of
hyperbolic equation (even for the largest time $\tau=100$). 

\subsection{Radiation boundary condition}

Here $b_1=-b_2>0$, $b_3<0$. The boundary condition takes the form of equation (\ref{eq27a})
where $\lambda=-b_1/b_3$. To obtain the inverse transforms of equations
(\ref{eq22}) and (\ref{eq23}) we assume that $\tau s\ll 1$ (what corresponds to $t\gg 1/\tau$) and we
extend the transforms into the power series with respect to the
parameter $s$. Achieving only the linear terms with respect to $\tau$ we
get
\begin{eqnarray}\label{eq28}
\fl C_1(x,t)=C_0-\frac{C_0}{2}\sum_{k=0}^{\infty}\left(-\frac{\sqrt{D_\alpha}}{2\lambda}\right)^k
\left[f_{k(1-\alpha/2)-1,\alpha/2}\left(t;\frac{-x}{\sqrt{D_\alpha}}\right)\right.\nonumber\\
\left.-\frac{k\tau}{2}f_{k(1-\alpha/2),\alpha/2}
\left(t;\frac{-x}{\sqrt{D_\alpha}}\right)
+\frac{x\tau}{2\sqrt{D_\alpha}}f_{k(1-\alpha/2)+\alpha/2,\alpha/2}\left(t;\frac{-x}{\sqrt{D_\alpha}}\right)\right],
\end{eqnarray}
\begin{eqnarray}\label{eq29}
\fl C_2(x,t)=\frac{C_0}{2}\sum_{k=0}^{\infty}\left(-\frac{\sqrt{D_\alpha}}{2\lambda}\right)^k
\left[f_{k(1-\alpha/2)-1,\alpha/2}\left(t;\frac{x}{\sqrt{D_\alpha}}\right)+\frac{k\tau}{2}f_{k(1-\alpha/2),\alpha/2}
\left(t;\frac{x}{\sqrt{D_\alpha}}\right)\right.\nonumber\\
-\left.\frac{x\tau}{2\sqrt{D_\alpha}}f_{k(1-\alpha/2)+\alpha/2,\alpha/2}\left(t;\frac{x}{\sqrt{D_\alpha}}\right)\right].
\end{eqnarray}

\begin{figure}[h]
		\centering
    \includegraphics{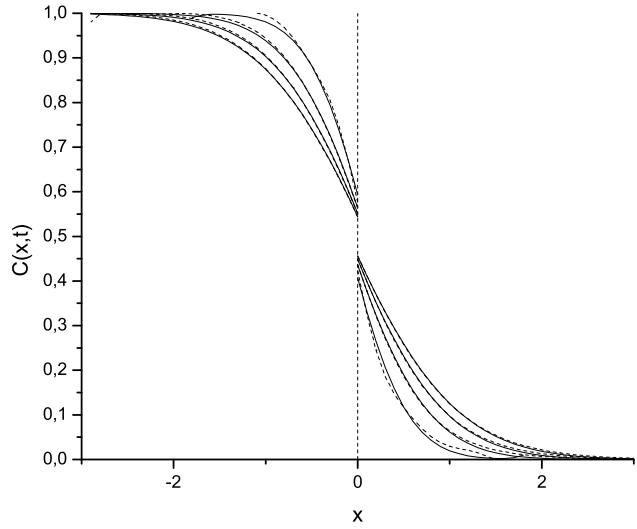}
    \caption{The solutions calculated for boundary conditions (\ref{5}) with $\lambda=10^{-3}$, the values of the parameters are the same as in \fref{fig:Fig5}.\label{fig:Fig6}}
\end{figure}

\begin{figure}[h]
		\centering
    \includegraphics{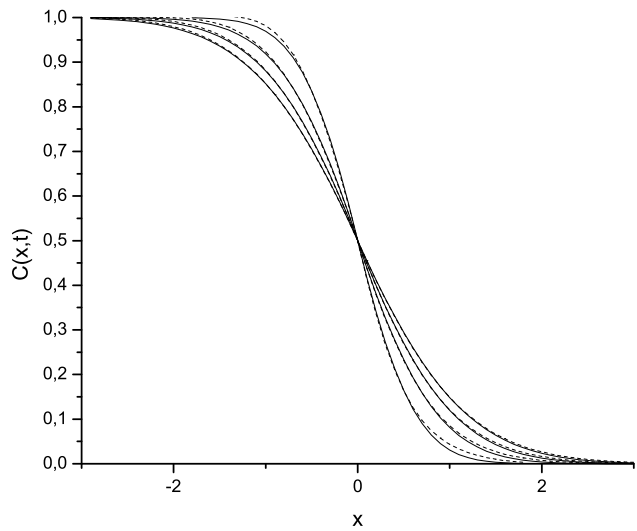}
    \caption{The solutions calculated for $\lambda=0.1$, the additional description is as in \fref{fig:Fig6}.\label{fig:Fig7}}
\end{figure}

The plots of functions (\ref{eq28}) and (\ref{eq29}) are presented in figures \ref{fig:Fig6} and \ref{fig:Fig7}. 

\section{Final remarks}

We present here the solutions of the parabolic and hyperbolic
subdiffusion equations for the homogeneous system and for the
membrane one. The solutions were found under assumption that we
take into account the terms linear with respect to the parameter
$\tau$. We applied two different boundary conditions at the membrane.
Our considerations are illustrated by few plots presenting the
solutions for both of the boundary conditions. The plots were
prepared for the parameters which values are of the order of the ones
already found for real systems on the basis of experimental results \cite{kdm}.
The detailed remarks extracted from the plots are not fully
conclusive, but it suggested few regularities, which - in our
opinion - are general. They are as follows.

\begin{enumerate}
    \item For the boundary condition (\ref{5})
\begin{itemize}
    \item The solutions at the membrane do not change in time and read as
    $C_1(0^-,t)=(\gamma C_0)/(1+\gamma)$, $C_2(0^+,t)=C_0/(1+\gamma)$.
    This property seems to be `unphysical', but let us note that the solution obtained for the system without
    membrane (for which $b_1=-b_2$ and $b_3=0$) with initial condition (\ref{eq21}) are constant for $x=0$ and reads as
    $C(0,t)=C_0/2$.
    \item The delay effect does not occur at the membrane, consequently it is weak at the membrane neighborhood.
\end{itemize}
    \item For radiation boundary condition (\ref{eq27a})
\begin{itemize}
    \item The concentration difference between the surfaces decreases in time (see \fref{fig:Fig6}).
    From equations (\ref{eq22}) and (\ref{eq23}) it is easy to see that $C_1(0,t)\rightarrow C_2(0,t)$ when $t\rightarrow \infty$, since the long time limit corresponds to the limit of small $s$.
    \item For $\lambda\sim 10^{-1}$ the membrane loses its selectivity.
\end{itemize}
		\item The main qualitative difference
between the above boundary conditions is noticeable in the long time limit as boundary condition
(\ref{5a}) leads to the solutions of hyperbolic subdiffusion equation which are going
to the continuous function at the barrier, unlike than the solutions obtained for (\ref{5}).
    \item In all cases the delayed effect is connected with the subdiffusion parameter $\alpha$
    (this property is clearly seen in \fref{fig:Fig1} - \fref{fig:Fig3} for the Green's functions).
    When $\alpha$ increases, the delaying effect is stronger.
    \item For large times the delaying effect is negligibly small. Let us note that for large time the term $\tau/t$ vanishes (it corresponds to the limit $\tau s\rightarrow 0$ in equations (\ref{eq22}) and (\ref{eq23})).
\end{enumerate}

Here the question arises: why the subdiffusion hyperbolic equation
has not been applied to describe the experimental results in
the subdiffusive membrane system, despite of proper `physical quality' of
the equation? Analyzing the plots \ref{fig:Fig2} - \ref{fig:Fig7} we conclude that in considered cases there is no
reason to apply the hyperbolic subdiffusion equation instead of
the parabolic one. The difference between the solutions is so
small that both of them would certainly be laid within the error
bars of the experimental concentration profiles. The order of values of the
subdiffusion coefficient $D_\alpha$ taken into calculations agrees
with the ones obtained experimentally for sugars in agarose gels
\cite{kdm} if as unit of time $1\;sec$ is chosen and $1\;mm$ is
the unit of space variable. In these units the value
$\tau=100$ is certainly too large, nevertheless these differences
are rather hard to observe, for smaller values of $\tau$ these
differences are smaller.

There is a problem with choosing the boundary condition at the
membrane. Seemingly there is no problem with choosing the condition
since the real system is limited by external walls and the
concentration goes to equilibrium functions which is continuous at
the membrane. This property posses the radiation boundary
condition only. However, experimental study performed on two membrane system show that the
concentration profiles have the scaling property, which is limited to the theoretical solutions obtained for the boundary conditions (\ref{5}) only \cite{dwor2006,koszt2008}. Namely, changing
the variables according to the relation $(x,t)\rightarrow (p^\alpha
x,pt)$, $p>0$ the experimental profiles does not display
noticeable changes. So, if the particles
flowing through the membrane do not `feel' the presence of
external walls of the system, the boundary conditions (\ref{5}) can be used.

Although our conclusion is rather odd in respect to the membrane
system, there are systems where the solutions of the hyperbolic
and parabolic equations considerably differ from each other. Such
a situation occurs for the boundary conditions where the
concentration of the particles oscillates with high frequency, as
for example in the problem of impedance spectroscopy \cite{lewkoszt}.

\ack
This paper was partially supported by
Polish Ministry of Education and Science under Grant No. 1 P03B 136
30.

\appendix
\section*{Appendix}
Here we proof the relation (\ref{eq10s}). Let us assume that $|C(x,u)|\leq A$ for $u\in(0,t]$ and $C(x,t)\rightarrow 0$ when $t\rightarrow 0$. Then,
\begin{displaymath}
	\left|\frac{\partial^{\alpha-1}C(x,t)}{\partial x^{\alpha-1}}\right|\leq\frac{A}{\Gamma(1-\alpha)}\int_0^t (t-u)^{-\alpha}du=\frac{A}{\Gamma(1-\alpha)}t^{1-\alpha} .
\end{displaymath}
From the above equation we get equation (\ref{eq10s}).

\end{document}